\documentstyle{article}
\newcommand{\bi}{\bibitem}

\input{amssym.def}

\begin{document}

\title{SUPERCOHERENT STATES OF CALOGERO-SUTHERLAND OSCILLATOR \\
}
\author{VLADISLAV G. BAGROV \\
{\it Department of Theoretical Physics, Tomsk State University, 36 Lenin Ave.%
} \\
{\it Tomsk, 634050, Russia } \and BORIS F. SAMSONOV \\
{\it Department of Theoretical Physics, Tomsk State University, 36 Lenin Ave.%
} \\
{\it Tomsk, 634050, Russia }}
\maketitle

\begin{abstract}
Supersymmetric quantum mechanical model of Calogero-Sutherlend singular
oscillator is constructed. Supercoherent states are defined with the help of
supergroup displacement operator. They are proper states of a fermionic
annihilation operator. Their coordinate and superholomorphic representations
are considered. The su\-per\-me\-a\-su\-re on superunit disc which realizes
the resolution of the unity is calculated. The cases of exact and
spontaneously broken supersymmetry are treated separately. As an example the
supersymmetric partners of the input Hamiltonian expressed in terms of
elementary functions are given.
\end{abstract}


\section{ Introduction}

The singular oscillator is a quantum mechanical system described by the
Hamiltonian
$$
h_0=-d^2/dx^2+x^2/4+b/x^2.\eqno (1)
$$
Several quantum mechanical problems and number of quantum field theory
problems are reduced to the Schr\"odinger equation with the Hamiltonian (1).
We can first cite the Coulomb problem and the problem of describing of
quantum system consisting of three (in general $N$) kinematically similar
particles in one dimension interacting pairwise via quadratic and
centrifugal potentials (Calogero problem \cite{i}). The Hamiltonian (1)
attracts now a considerable interest in connection with its usage for
describing of spin chains \cite{ii,iii}, quantum Hall effect \cite{iv},
fractional statistics and anyons \cite{v,vi}.

The Hamiltonian (1) has only discrete spectrum on the half-line $[0,\infty )$%
. The systems of coherent states for this Hamiltonian are well known and
studied (see for example \cite{vii}). These states are defined with the help
of the displacement operator for the group SU(1.1) which is a dynamical
symmetry group for the system under consideration and represent a continuous
basis in the Hilbert space of functions square integrable on the half-line.
Coherent states can also be defined using ladder operator as the eigenstates
of the destruction operator. The difference of these two definitions was
recently analyzed in \cite{viii}.

It is known that using the Darboux transformation one can construct the
super\-ha\-mil\-to\-ni\-an (se, for example, \cite{ix}) of the Witten
supersymmetric quantum mechanics \cite{x}. One can find the review of the
results obtained in supersymmetric quantum mechanics in papers \cite{ufn,xi}%
. This method was recently generalized to a time-dependent Schr\"odinger
equation \cite{pl,jp,j} and applied to constructing of coherent states of
anharmonic oscillator Hamiltonians with equidistant and quaziequidistant
spectra \cite{jp,j} and coherent states of the Hamiltonian of the one
soliton potential \cite{np}.

The notion of coherent states is generalized in su\-per\-sym\-met\-ric
quantum mechanics up to supercoherent ones. Supercoherent states for
su\-per\-sym\-met\-ric harmonic oscillator have been introduced in \cite
{xiv,xv} and then generalized to arbitrary Lie supergroups \cite{xvi} in the
frame of Perelomov's coherent states \cite{vii}. These states defined with
the help of the supergroup displacement operator was recently applied for
the description of coherent states of spinning particle in varying
electromagnetic field \cite{xvii}. We will mention as well the papers \cite
{xviii,xix} in which the detailed analysis of supercoherent states and their
underlying geometric structures for the supergroups $OSP(1/2)$ and $OSP(2/2)$
have been made.

In this paper we use the Darboux transformation operator to construct the
su\-per\-sym\-met\-ric partner of the Hamiltonian $h_0$. Using this
transformation we obtain two copies of the Hilbert space with a positive
definite quadratic form (scalar product) and a representation of the algebra
$su(1.1)$ defined on them. These two Hilbert spaces are essential elements
in constructing of the coordinate representation of a linear superspace
which becomes the Hilbert superspace after definition on it a superscalar
product. Then one defines superalgebra generators which form a coordinate
representation of the dynamical superalgebra of the system with given
superhamiltonian. The supercoherent states are defined with the help of the
sipergroup displacement operator. Their coordinate representation is
obtained. The representation of these states in the space of the functions
superholomorphic in a superunit disc is considered. Supermeasure which
realizes resolution of the identity operator in the given superspace is
calculated. The possibility of the spontaneous supersymmetry breaking down
is taken into account. In conclusion concrete examples of transformations
which give exact and spontaneously broken supersymmetry are given.

\section{ Coherent States of Singular Oscillator}

We shall give now a brief survey of the well known \cite{vii} properties of
coherent states of the singular oscillator. The dynamical symmetry algebra
for the system with Hamiltonian (1) is $su(1.1)$. Its Cartan-Weil basis in
coordinate representation is expressed via the harmonic oscillator
annihilation $a=d/dx+x/2$ and creation $a^{+}=-d/dx+x/2$ operators as
follows
$$
k_0=(1/2)h_0,\quad k_{+}=(1/2)\left[ (a^{+})^2-b/x^2\right] ,\quad
k_{-}=(1/2)\left[ a^2-b/x^2\right] ,
$$
$$
[k_0,k_{\pm }]=\pm k_{\pm },\quad [k_{-},k_{+}]=2k_0.
$$
Casimir operator in this representation takes a constant value ${\cal C}%
=\frac 12\left[ k_{+}k_{-}+k_{-}k_{+}\right] -k_0^2=3/16-b/4$. Discrete
representation of $su(1.1)$ algebra corresponding to the value of $%
k=1/2+(1/4)\sqrt{1+4b}=\frac 12E_0$, ($E_0$ being the ground state energy)
is defined by the conditions $k_{-}\mid 0\rangle =0$, $k_0\mid 0\rangle
=k\mid 0\rangle $, $k_{+}~\mid ~n~\rangle =\sqrt{(n+1)(n+2k)}~\mid
~n~+~1~\rangle $. Let $H_0$ be the Hilbert space of the solutions of the
time-dependent Schr\"odinger equation with the Hamiltonian (1) and $\mid
n,t\rangle $ be its discrete basis vectors. In coordinate representation
every function $\psi _n(x,t)=\langle x\mid n,t\rangle $ satisfies the zero
boundary condition on half-line $[0,\infty )$. We shall consider the
functions $\mid n\rangle =\mid n,0\rangle $ defined by the action of the
raising operator $k_{+}$:
$$
\mid n\rangle =\left[ (n!)^{-1}\Gamma ^{-1}(2k+n)\Gamma (2k)\right]
^{1/2}\left( k_{+}\right) ^n\mid 0\rangle ,\quad k_0\mid n\rangle =(k+n)\mid
n\rangle .
$$
Their coordinate representation is as follows:
$$
\psi _n(x)=\left[ n!2^{1-2k}\Gamma ^{-1}(n+2k)\right] ^{1/2}x^{2k-1/2}\exp
(-x^2/4)L_n^{2k-1}(x^2/2),\eqno (2)
$$
where $L_n^\alpha (z)$ is the Laguerre polynomial.

Coherent states $\mid z\rangle $ are defined by the action of the
displacement operator $D_z$ for the group $SU(1.1)$
$$
D_z=e^{zk_{+}}\exp \left[ \ln (1-z\overline{z})k_0\right] e^{-\overline{z}%
k_{-}},\quad z\in {\Bbb C},\quad \left| z\right| <1
$$
on the vacuum state%
$$
\mid z\rangle =D_z\mid 0\rangle =(1-z\overline{z})^k\sum\nolimits_{n=0}^%
\infty \left[ (n!)^{-1}\Gamma ^{-1}(2k)\Gamma (2k+n)\right] ^{1/2}z^n\mid
n\rangle .
$$
In the coordinate representation we have
$$
\psi _z(x)=\langle x\mid z\rangle =2^{1/2-k}\Gamma ^{-1/2}(2k)(1-z)^{-2k}(1-z%
\overline{z})^kx^{2k-1/2}\exp \left[ {\textstyle {\frac{-(1+z)x^2 }{4(1-z)}}}%
\right] .
\eqno (3)
$$
As $\overline{z}$ we denote the value complex conjugated to $z$.

The functions $\psi _z(x)$ realize in the space $H_0$ an overcomplete basis
set far which the resolution of the unity reads as follows
$$
\int\nolimits_{\left| z\right| <1}\mid z\rangle \langle z\mid d\mu
(z)=1,\quad d\mu (z)=(1/\pi )(2k-1)(1-z\overline{z})^{-2}dzd\overline{z}.
\eqno (4)%
$$

The decomposition coefficients $C_n$ of any $\mid \psi \rangle \in H_0$ in
terms of the discrete basis $\mid n\rangle $, $\mid \psi \rangle
=\sum\limits_{n=0}^\infty C_n\mid n\rangle $ define its holomorphic
representation $\psi (z)$ in the space of functions holomorphic in the unit
disk $\mid z\mid <1$. Scalar product in this space is defined as follows
$$
\langle \psi _1(z)\mid \psi _2(z)\rangle =\int\nolimits_{\left| z\right|
<1}e^{-f}\psi _1(\overline{z})\psi _2(z)d\mu (z), \eqno (5)%
$$
where $f=\ln \left| \langle 0\mid z\rangle \right| ^{-2}=\ln (1-z\overline{z}%
)^{-2k}$.

\section{Supersymmetry of Singular Oscillator}

Following the papers \cite{ix,xx,xxi} we construct a new exactly solvable
potential $h_1=h_0+A(x)$ with the help of the Darboux transformation which
is defined by a nodeless in half-line $(0,\infty )$ real solution $u(x)$ of
the initial Schr\"odinger equation $h_0u=\alpha u$ called {\it %
transformation function}, where $\alpha (<2k)$ is an arbitrary constant. The
transformation function $u(x)$ completely defines the potential difference
$$
A(x)=-2\left[ \ln u(x)\right] ^{\prime \prime },\eqno (6)
$$
where the prime denotes the derivative with respect to $x$ and {\it %
transformation operator} is defined as follows: $\widetilde{L}=-u^{\prime
}(x)/u(x)+d/dx$. Operator $\widetilde{L}$ transforms the solutions $\psi (x)$
of the input Schr\"odinger equation
$$
h_0\psi (x)=E\psi (x)\eqno (7)
$$
into the solutions of the transformed one
$$
h_1\varphi (x)=E\varphi (x),\quad \varphi (x)=\widetilde{L}\psi (x)\eqno (8)
$$
corresponding to the same eigenvalue $E$. The operator $\widetilde{L}^{+}$
adjoint to $\widetilde{L}$ with respect to scalar product for which $%
(d/dx)^{+}=-d/dx$ realizes the transformation in the inverse direction, i.e.
from the solutions $\varphi (x)$ of the equation (8) to the solutions $\psi
(x)$ of the equation (7). The products of the operators $\widetilde{L}$ and $%
\widetilde{L}^{+}$ are symmetry operators for the equations (7) and (8) and
are expressed through the Hamiltonians $h_0$ and $h_1$: $\widetilde{L}^{+}%
\widetilde{L}=h_0-\alpha $, $\widetilde{L}\widetilde{L}^{+}=h_1-\alpha $.
Moreover the operators $\widetilde{L}^{+}$ and $\widetilde{L}$ intertwine
the Hamiltonians $h_0$ and $h_1$
$$
\widetilde{L}h_0=h_1\widetilde{L},\quad h_0\widetilde{L}^{+}=\widetilde{L}%
^{+}h_1.
$$
If the functions $\psi (x)=\psi _n(x)$ with eigenvalues $E_n$ are normalized
to unity we easily obtain the norm of the functions $\varphi _n(x)=%
\widetilde{L}\psi _n(x)$: $\langle \varphi _n\mid \varphi _n\rangle
=E_n-\alpha $.

If the function $u$ becomes infinity on the bounds of the interval $%
[0,\infty )$ then the equation (8) has the following function $\varphi
_{-1}(x)=u^{-1}(x)\in H_1$ as its solution. It is square integrable on this
interval and satisfies the boundary condition for the discrete spectrum
eigenfunctions of the equation (8). We denote as $H_1$ a Hilbert space of
the solutions of the time-dependent Schr\"odinger equation with the
Hamiltonian $h_1$, so $\varphi _n(x)=\varphi _n(x,0)$, $\varphi _n(x,t)\in
H_1$. There exist in $H_0$ such functions $\psi _n(x)$ that $L\psi _n(x)\in
H_1$. The function $\varphi _{-1}(x)$ will be the ground state function of
the Hamiltonian $h_1$ and we shall obtain the exact supersymmetry \cite
{x,ufn}. We shall as well consider the case when $u(0)=0$ and $u(\infty
)=\infty $ so that $u^{-1}(x)\notin H_1$. In this case we obtain the
spontaneously broken supersymmetry. It should be noted that the functions $u$
and $\varphi _{-1}(x)$ form the kernels of the operators $\widetilde{L}$ and
$\widetilde{L}^{+}$: $\widetilde{L}u=0$, $\widetilde{L}^{+}\varphi
_{-1}(x)=0 $. The case $u=\psi _0(x)$ reduces to the considered one after
the replacement $h_0\longleftrightarrow h_1$.

We will use the means of the supermathematics \cite{xxii,xxiii,xxiv}
together with the transformation operators
$$
L=\widetilde{L}\left( h_0-\alpha \right) ^{-1/2}=\left( h_1-\alpha \right)
^{- 1/2}\widetilde{L} \eqno (9)%
$$
and its conjugate
$$
L^{+}=\widetilde{L}^{+}\left( h_1-\alpha \right) ^{-1/2}=\left( h_0-\alpha
\right) ^{-1/2}\widetilde{L}^{+}.%
$$
These operators are such that $LL^{+}=1$ and $L^{+}L=1$ in the corresponding
spaces. If the supersymmetry is broken $L\psi\neq 0$, $\forall\psi\in H_0$, $%
L^+\varphi\neq 0$, $\forall\varphi\in H_1$ and also $h_0-\alpha \neq 0$, $%
h_1-\alpha \neq 0$ in $H_0$ and $H_1$ respectively. Operators $L$ and $L^+$
are in this case unitary operators and consequently they transform one
orthonormal basis set into another one.

If supersymmetry is exact ($\varphi _{-1}=u^{-1}\in H_1$) we can decompose
the space $H_1^1$ into a direct sum $H_1=H_1^0\bigoplus H_1^1$ where $H_1^0=$
span $\left\{ \varphi _{-1}\right\} $ and $H_1^1=$ span $\left\{ \varphi
_n=L\psi _n,\forall \psi _n\in H_0\right\} $. The symbol span stands for the
linear hull over the complex number field ${\Bbb C}$. The relations $%
LL^{+}=1 $ and $h_1-\alpha \neq 0$ are valid for any $\varphi \in H_1^1$.

We shall construct the linear superspace $H_s$ with the help of the spaces $%
H_0$ and $H_1$ and the generators $\theta $ and $\overline{\theta }$ of the
four-dimensional Grassman algebra: $\overline{\overline{\theta }}=\theta $, $%
\theta ^2=\overline{\theta }^2=\theta \overline{\theta }+\overline{\theta }%
\theta =0$ using so-called homogeneous realization: $H_s=H_{\overline{0}%
}\oplus H_{\overline{1}}$. The even subspace $H_{\overline{0}}$ is defined
as follows $H_{\overline{0}}=$span$\left\{ \Psi _n^1(x,\theta ,\overline{%
\theta })=\psi _n(x),n=0,1,2,\ldots \right\} $. When one defines the odd
component $H_{\overline{1}}$ it is necessary to distinguish the cases of
broken and exact supersymmetry. For spontaneously broken supersymmetry the
space $H_{\overline{1}}$ is defined as follows $H_{\overline{1}}=$span$%
\left\{ \Psi _n^1(x,\theta ,\overline{\theta })=\theta \varphi
_n(x),n=0,1,2,\ldots \right\} $, $\varphi _n(x)=L\psi _n(x)$. If the
supersymmetry is exact the same linear hull defines the space $H_{\overline{1%
}}^1$. The space $H_{\overline{1}}$ is in this case a direct sum $H_{%
\overline{1}}=H_{\overline{1}}^0\oplus H_{\overline{1}}^1$, where $H_{%
\overline{1}}^0=$span$\left\{ \theta \varphi _{-1}(x)\right\} $. The whole
space $H_s$ is in the latter case as well a direct sum $H_s=H_s^1\oplus H_{%
\overline{1}}^0$, where $H_s^1=H_{\overline{0}}\oplus H_{\overline{1}}^1$.
The elements of the spaces $H_{\overline{0}}$ and $H_{\overline{1}}$ are
called {\it homogeneous elements}. For every homogeneous element $v\in H_s$
one defines its parity $\varepsilon (v)$ as follows: $\varepsilon (v)=0$ if $%
v\in H_{\overline{0}}$ and $\varepsilon (v)=1$ if $v\in H_{\overline{1}}$.

For the purpose of construction of coherent states we need the notion of the
Grassmann envelop of the second kind. To define this notion we shall
consider the four-dimensional Grassmann algebra $\Lambda _2$ with the
generators $\alpha $ and $\overline{\alpha }$ ($\overline{\overline{\alpha }}%
=\alpha $, $\alpha ^2=\overline{\alpha }^2=\alpha \overline{\alpha }+%
\overline{\alpha }\alpha =0$). We define the Grassmann envelop of the second
kind $\widetilde{H}_s$ of the superspace $H_s$ as a left $\Lambda _2$-
module \cite{xxii}: $\widetilde{H}_s=\left( \Lambda _2\otimes H_s\right) _{%
\overline{0}}=\left( \Lambda _2^0\otimes H_{\overline{0}}\right) \oplus
\left( \Lambda _2^1\otimes H_{\overline{1}}\right) $. Here $\Lambda _2^0$ is
the commutative subalgebra of the even elements of the algebra $\Lambda _2$ (%
$c$ - numbers), $\Lambda _2^1$ is the space of the odd elements ($a$ -
numbers) and $\Lambda _2=\Lambda _2^0\oplus \Lambda _2^1$ (supernumbers)
\cite{xxiv}. We notice the following property: $\beta v=(-1)^{\varepsilon
(\beta )\varepsilon (v)}v\beta $ which holds for homogeneous elements $v\in
H_s$ and $\beta \in \Lambda _2$.

To define the space $\overline{H}_s$ dual to $H_s$ and the module $\overline{%
\widetilde{H}}_s$ dual to $\widetilde{H}_s$ we need the definition of the
complex conjugation. We define the complex conjugation by the following
property
$$
\overline{\beta v}=\overline{\beta }\overline{v},\quad \forall v\in
H_s,\quad \forall \beta \in \Lambda _2.\eqno (10)
$$
This definition differs from the widely used condition $\overline{\beta
_1\beta _2}=\overline{\beta }_2\overline{\beta }_1$, $\beta _{1,2}\in
\Lambda _2$ \cite{xxii,xxiii,xxiv}. In the latter case the quantity $\beta
\overline{\beta }$ is real independently of the parity of an element $\beta
\in \Lambda _2$. We follow the definition (10) since in the other case one
faces a number of inconsistencies. In particular, the super K\"ahler 2-form
is neither real nor imaginary \cite{xix} and it is difficult to establish in
this case the correspondence between the real observables and self
superadjoint operators.

The definition of the space $\widetilde{H}_s$ dual to $H_s$ is equivalent to
the definition of the bilinear form on $H_s$ with respect to which the space
$\widetilde{H}_s$ will be the space of the linear functionals defined on $%
H_s $. We define first the dual space in such a way that bilinear form on $%
H_s$ be expressed in terms of bilinear forms (scalar products) defined in $%
H_0$ and $H_1$ in accordance with the following property \cite{xix}
$$
\langle v_1\mid v_2\rangle =\langle \psi _1\mid \psi _2\rangle +i\langle
\varphi _1\mid \varphi _2\rangle ,
$$
where $v_1=\psi _1+\theta \varphi _1$, $v_2=\psi _2+\theta \varphi _2$ are
elements of the space $H_s$ in their homogeneous realization and $\langle
\psi _1\mid \psi _2\rangle $, $\langle \varphi _1\mid \varphi _2\rangle $
are the usual scalar products defined in the spaces $H_0$ and $H_1$. For
this purpose consider the homogeneous decomposition of the dual space $%
\overline{H}_s=\overline{H}_{\overline{0}}\oplus \overline{H}_{\overline{1}}$%
. We suppose that if we pass from $H_s$ to $\overline{H}_s$ the parity of
the elements does not change. This condition can be realized if with every
element $\Psi _n^0(x,\theta ,\overline{\theta })=\psi _n\in H_{\overline{0}}$
one associates an element $\overline{\Psi }_n^0(x,\theta ,\overline{\theta }%
)=\overline{\theta }\theta \overline{\psi }_n\in \overline{H}_{\overline{0}}$
and with every element $\Psi _n^1(x,\theta ,\overline{\theta })=\theta
\varphi _n\in H_{\overline{1}}$ one associates an element $\overline{\Psi }%
_n^1(x,\theta ,\overline{\theta })=\overline{\theta }\overline{\varphi }%
_n\in \overline{H}_{\overline{1}}$. For spontaneously broken supersymmetry
the spaces $\overline{H}_{\overline{0}}$ and $\overline{H}_{\overline{1}}$
are defined as the linear hulls over the complex number field ${\Bbb C}$ of
the vectors $\overline{\Psi }_n^0$ and $\overline{\Psi }_n^1$, $%
n=0,1,2,\ldots $ respectively. For exact supersymmetry we shall have the
direct sum $\overline{H}_{\overline{1}}=\overline{H}_{\overline{1}}^0\oplus
\overline{H}_{\overline{1}}^1$, where $\overline{H}_{\overline{1}}^0=$span$%
\left\{ \overline{\Psi }_{-1}=\overline{\theta }\overline{\varphi }%
_{-1}(x)\right\} $ and $\overline{H}_{\overline{1}}^1=$span$\left\{
\overline{\Psi }_n^1,\ \ n=0,1,2,\ldots \right\} $. When the space $H_s$ is
considered as left $\lambda _2$ - module one has to take into account the
property (10).

Since we have the one-to-one correspondence between the elements of the
spaces $H_s$ and $\overline{H}_s$ we can define the bilinear form $\langle
\cdot \mid \cdot \rangle $: $H_s\otimes H_s\rightarrow {\Bbb C}$ as follows
$$
\langle \Psi _1\mid \Psi _2\rangle =\int \overline{\Psi }_1(x,\theta ,%
\overline{\theta })\Psi _2(x,\theta ,\overline{\theta })dxd\theta d\overline{%
\theta }\in {\Bbb C} . \eqno (11)%
$$
The integration over the Grassmann variables is understood in the sense of
Berezin \cite{xxii} when the only non-zero integral is $\int \overline{%
\theta }\theta d\theta d\overline{\theta }=1$. When $H_s$ is considered as
left $\Lambda _2$ - module this definition realizes the following
correspondence: $\widetilde{H}_s\otimes \widetilde{H}_s\rightarrow \Lambda
_2^0$. We notice the property $\overline{\langle v\mid w\rangle }%
=(-1)^{\varepsilon (v)\varepsilon (w)}\langle w\mid v\rangle $ which holds
for homogeneous elements $v,w\in H_s$. One should work with supernumbers
according to the following rule: $\langle \beta _1v\mid \beta _2w\rangle
=(-1)^{\varepsilon (v)\varepsilon (\beta _2)} \overline{\beta }_1\beta _2
\langle v\mid w\rangle $, where $\beta _2$ and $v$ are homogeneous elements
of $\Lambda _2$ and $H_s$ respectively.

The comparison of the bilinear form (11) with the super Hermitian form
defined in an abstract supervector space \cite{xix} shows that we have a
concrete (coordinate) realization of the abstract super Hilbert space with
(11) as the super Hermitian form. It is worth noticing that there exist
other definitions of super Hilbert spaces, see for example \cite{HSS}.

We pass now to definition of the coordinate representation of generators of
the superalgebra acting in the space $H_s$. Let us put
$$
K_0=k_0(\partial /\partial \theta )\theta +Lk_0L^{+}\theta (\partial
/\partial \theta ),\quad K_{\pm }=k_{\pm }(\partial /\partial \theta )\theta
+Lk_{\pm }L^{+}\theta (\partial /\partial \theta ).\eqno (12)
$$
Here $\theta $ is an operator of the left multiplication on the element $%
\theta $ and $\frac \partial {\partial \theta }$ is an operator of the left
differentiation. (We define the left action of the operators on the vectors $%
\Psi (x,\theta ,\overline{\theta })\in H_s$). Operators (12) realize
evidently the coordinate representation of the algebra $su(1.1)$ in the
space $H_s$. The identity operator in this space has the form $I=\frac
\partial {\partial \theta }\theta +\theta \frac \partial {\partial \theta }$%
. We note as well another even operator: $B_0=\frac \partial {\partial
\theta }\theta -\theta \frac \partial {\partial \theta }$. It has the
property $B_0v=(-1)^{\varepsilon (v)}v$ for all homogeneous $v\in H_s$.

All the operators introduced do not change the parity of the elements and
consequently they are even operators. We define odd operators by the
relations
$$
Q_{-}=L^{+}(\partial /\partial \theta ),\quad Q_{+}=L\theta .\eqno (13)
$$
They commute with all even operators and $\left\{ Q_{-},Q_{+}\right\} =I$.
It follows from this that the linear hull $sal=$span$\left\{
K_0,K_{+},K_{-},I,Q_{+},Q_{-}\right\} $ represents a superalgebra. For this
superalgebra we can write the following direct sums $sal=sal_{\overline{0}%
}\oplus sal_{\overline{1}}$, $sal_{\overline{0}}=su(1.1)_{\overline{0}%
}\oplus e_{\overline{0}}$, where $su(1.1)_{\overline{0}}=$span$\left\{
K_0,K_{+},K_{-}\right\} $, $e_{\overline{0}}=$span$\left\{ I\right\} $ and $%
sal_{\overline{1}}=$span$\left\{ Q_{+},Q_{-}\right\} $.

Having in hand the super Hermitian form (11) on $H_s$ we define operator $%
A^{+}$ super adjoint to $A$ \cite{xix}.
$$
\langle A^{+}v\mid w\rangle =\left( -1\right) ^{\varepsilon (v)\varepsilon
(A)}\langle v\mid Aw\rangle ,\quad \forall v,w\in H_s,\quad A\in sal
$$
$v$ and $A$ being homogeneous elements of $H_s$ and $sal$ respectively. The
conjugation operation so defined has the properties $(A^{+})^{+}=A$, $%
(AB)^{+}=(-1)^{\varepsilon (A)\varepsilon (B)}B^{+}A^{+}$ and $\left[
A,B\right] ^{+}=-\left[ A^{+},B^{+}\right] $, where $\left[ ,\right] $ is
superalgebra bracket \cite{xxii} with the property \\ $\left[ \beta
_1A,\beta _2B\right] =(-1)^{\varepsilon (A)\varepsilon (\beta _2)}\beta
_1\beta _2\left[ A,B\right] $. Self super adjoint operator is defined as
follows $A^{+}=A$. It is easily seeing that $K_0^{+}=K_0$, $K_{\pm
}^{+}=K_{\mp }$, $I^{+}=I$, $B_0^{+}=B_0$, $Q_{\pm }^{+}=iQ_{\mp }$.
Operator ${\cal H}=2K_0$ can be treated as superhamiltonian since ${\cal H}%
^{+}={\cal H}$ and the spectral problem for it considered in the spaces $H_{%
\overline{0}}$ and $H_{\overline{1}}$ is reduced to the solution of the
input and output Schr\"odinger equations. Superalgebra $sal$ is evidently
its dynamical supersymmetry algebra. It is worth noticing that the operators
(12), (13) and $I$ in the case of spontaneously broken supersymmetry realize
irreducible representation of this superalgebra in the space $H_s$. If the
supersymmetry is exact this representation is reducible one and the direct
decomposition $H_s=H_s^1\oplus H_{\overline{1}}^0$ represents the
decomposition of this representation in terms of irreducible ones. The
representation of the superalgebra $sal$ in the space $H_{\overline{1}}^0$
is trivial: $gu=0$, $\forall g\in sal$, $\forall u\in H_{\overline{1}}^0$.
The whole spectrum of the superhamiltonian ${\cal H}$ is two-fold degenerate
in the case of the spontaneously broken supersymmetry. In the case of the
exact supersymmetry the ground state level $E=\alpha $ of the
superhamiltonian ${\cal H}$ is nondegenerate and the vacuum state is
described by the function $\Psi _{-1}(x,\theta ,\overline{\theta })=\theta
\varphi _{-1}(x)$. If we define superunitary operator by the condition $%
U^{-1}=U^{+}$ we can construct superunitary symmetry operators which form
local supergroup with superalgebra $sal$.

We need the completeness condition of the basis set $\left\{ \Psi _n^0,\Psi
_n^1,n=0,1,\ldots \right\} $ in the space $H_s$ (in $H_s^1$ for the exact
supersymmetry). To formulate this condition we define the projection
operator $P_v$: $P_vw=v\langle v\mid w\rangle $, where the superscalar
product is defined by the formula (11). Let us introduce the notations: $%
P_n^0=P_{\Psi _n^0}$ $P_n^1=P_{\Psi _n^1}$. It is evident that $%
P_n^0P_{n^{\prime }}^1=0$, $\forall n,n^{\prime }$ and operator $%
P_n=P_n^0-iP_n^1$ projects on the two-dimensional space of the solutions of
the super Schr\"odinger equation ${\cal H}\Psi =E\Psi $ with given value of $%
E$. The completeness condition of the basis $\left\{ \Psi _n^0,\Psi
_n^1\right\} $ reads as follows: $\sum_{n=0}^\infty P_n=I$.

\section{ Supercoherent Stattes}

We define the supercoherent states using supergroup displacement operator
\cite{xvi,xvii,xviii}.

The state $\Psi _0^0$ is of maximal symmetry state since it is the proper
state of the operators $K_0=\frac 12{\cal H}$, $K_{-}$, $Q_{-}$, and $I$
forming the algebra ${\cal B}$ of the isotropy subgroup of the element $\Psi
_0^0$. In this case the complexification of $sal$, $sal^c$, is decomposed in
the direct sum $sal^c={\cal B}\oplus \overline{{\cal B}}$, where $\overline{%
{\cal B}}=$span$\left\{ K_0,I,K_{+},Q_{+}\right\} $. This is the reason to
take $\Psi _0^0$ as a cyclic vector (fiducial state) and apply to it the
nonunitary supergroup translation operator.
$$
\Psi _{z\alpha }(x,\theta ,\overline{\theta })=N\exp \left( zK_{+}-\alpha
Q_{+}\right) \Psi _0^0,\quad z\in {\Bbb C},\quad \alpha \in \Lambda _2.
$$
The normalizing coefficient $N$ should be calculated separately. Taking into
account the commutativity of $Q_{+}$ and $K_{+}$ one finds
$$
\Psi _{z\alpha }(x,\theta ,\overline{\theta })=N\left( \psi _z(x)-\alpha
\theta \varphi _z(x)\right) ,\quad \varphi _z(x)=L\psi _z(x).\eqno (14)
$$
The direct calculation shows that this function is a proper function of
(fermionic) annihilation operator: $Q_{-}\Psi _{z\alpha }=\alpha \Psi
_{z\alpha }$.

Using the fact that $\psi _z(x)$ is normalized to unity and with the help of
the formula (11) we find the normalization coefficient $N=1+\frac 12i%
\overline{\alpha }\alpha =\overline{N}$.

It is worth noticing that in the paper \cite{xxv} the analogous expression
for coherent states of the harmonic oscillator potential has been obtained.

Using the completeness of the basis $\left\{ \Psi _n^0,\Psi _n^1\right\} $
we can find the su\-per\-me\-a\-su\-re\\ $d\mu (z,\overline{z},\alpha ,%
\overline{\alpha })=I$ which gives the resolution of the identity in the
space $H_s$
$$
\int\nolimits_{{\cal D}}P_{\overline{z}\overline{\alpha }}d\mu (z,\overline{z%
},\alpha ,\overline{\alpha })=I
$$
where $P_{\overline{z}\overline{\alpha }}$ is the projection operator on the
state $\Psi _{\overline{z}\overline{\alpha }}(x,\theta ,\overline{\theta })$
and ${\cal D}$ is a domain in the superspace with coordinates $(z,\alpha )$,
$\mid z\mid <1$, called {\it super unit disc}.

A simple computation leads to
$$
d\mu (z,\overline{z},\alpha ,\overline{\alpha })=id\alpha d\overline{\alpha }%
d\mu (z,\overline{z}),
$$
where $d\mu (z,\overline{z})$ is determined by the formula (4).

The resolution of the identity permits one to construct a superholomorphic
representation of the vectors from $H_s$. For this purpose we write an
element $\Psi (x,\theta ,\overline{\theta })\in H_s$ as follows
$$
\Psi (x,\theta ,\overline{\theta })=\sum\nolimits_{n=0}^\infty \left[
C_n^0\psi _n(x)+C_n^1\theta \varphi _n(x)\right] =\psi (x)+\theta \varphi
(x).
$$
It follows from this relation that
$$
\langle \Psi _{\overline{z}\overline{\alpha }}\mid \Psi \rangle =\left( 1-z%
\overline{z}\right) ^k\left( 1-{\textstyle {\frac{ }{}}}\overline{\alpha }%
\alpha \right) \Psi (z,\alpha ),\quad \Psi (z,\alpha )=\psi (z)-i\alpha
\varphi (z),
$$
where $\psi (z)$ and $\varphi (z)$ are holomorphic representations of
functions $\psi (x)$ and $\varphi (x)$. Superfunction $\Psi (z,\alpha )$ is
the holomorphic representation of the vector $\Psi $.

In the space of superholomorphic functions we can define the scalar product
$$
\langle \Psi _1(z,\alpha )\mid \Psi _2(z,\alpha )\rangle =\int\nolimits_{}D%
\overline{\Psi }_1(z,\alpha )\Psi _2(z,\alpha )e^{-f}d\mu , \eqno (15)
$$
where superfunction $f$ is defined by the condition \cite{xix} $f=f(z,%
\overline{z},\alpha ,\overline{\alpha })=\ln \left| \langle \Psi _0^0\mid
\Psi _{\overline{z}\overline{\alpha }}\rangle \right| ^{-2}$ so that $\exp
\left( -f\right) =\left( 1-i\alpha \overline{\alpha }\right) \left( 1-z%
\overline{z}\right) ^{2k}$. The formula (15) leads to the same definition
for the scalar product: $\langle \Psi _1\mid \Psi _2\rangle =\langle \psi
_1\mid \psi _2\rangle +i\langle \varphi _1\mid \varphi _2\rangle $, where $%
\langle \psi _1\mid \psi _2\rangle $ and $\langle \varphi _1\mid \varphi
_2\rangle $ are scalar products defined for the holomorphic representations
of the vectors $\psi $ and $\varphi $.

\section{Example}

As an example consider a solution of the elementary form of the
Schr\"odinger equation with the Hamiltonian (1) as the transformation
functions. The Hamiltonian $h_1$ will be expressed in this case in terms of
elementary functions.

Singular at the bounds of the interval $[0,\infty )$ solutions of the
Schr\"odinger equation with Hamiltonian (1) can be chosen as follows
$$
u_p(x)=x^{3/2-2k}e^{x^2/4}L_p^{1-2k}(y),\ \ y=-x^2/2, h_0u_p=2(k-p-1)u_p,\ \
p=0,1,2,\ldots .\eqno (16)
$$
With the help of the formula (6) we find out the potential difference
$$
A_p(x)=-1+\frac{3-4k}{x^2}+2\left[ \frac{xL_{p-1}^{2-2k}(y)}{L_p^{1-2k}(y)}%
\right] ^2-2\frac{x^2L_{p-2}^{3-2k}(y)+L_{p-1}^{2-2k}(y)}{L_p^{1-2k}(y)}.%
\eqno (17)
$$
The transformation operator which transforms the solutions (2) of the input
equation (7) into the solutions of the equation (8) reads as follows
$$
L=\left[ \frac{4k-3}{2x}-\frac x2-\frac{xL_{p-1}^{2-2k}(y)}{L_p^{1-2k}(y)}%
-\frac d{dx}\right] \left( h_0+2+2p-2k\right) ^{-1/2}.\eqno (18)
$$
The solution of the equation (7) which can not be obtained by the action of
the operator (18) to the functions from $H_0$ is $u_p^{-1}(x)$. This
function square integrable on half-line $[0,\infty )$ only at even $p$. Its
normalization constant can be calculated by the method described in \cite
{xxvi}. One then obtains
$$
\int\nolimits_0^\infty u_p^{-2}(x)dx=(-1)^p2^{2k-2}p!\Gamma (2k-p-1).
$$
At odd $p$ the potential differences has poles on half-line $[0,\infty )$.
At even $p$ transformation function (16) generates the exact supersymmetry
with vacuum state constructed using the normalized at unity ground state of
the new Hamiltonian
$$
\varphi _{-1}(x)=2^{k-1}\sqrt{p!\Gamma (2k-p-1)}u_p^{-1}(x).
$$

To construct the spontaneously broken supersymmetry we use the
transformation function which vanishes at $x=0$ and equals infinity at $%
x=\infty $
$$
v_p(x)=x^{2k-1/2}e^{x^2/4}L_p^{2k-1}(y),\quad h_0v_p=-2(k+p)v_p,\quad
p=0,1,2,\ldots .
$$
This function is not square integrable on half-line $[0,\infty )$ (the
normalization integral diverges at $x=0$), but the potential difference
calculated with it
$$
A_p(x)=-1+\frac{4k-1}{x^2}+2x^2\left[ \frac{L_{p-1}^{2k}(y)}{L_p^{2k-1}(y)}%
\right] ^2-2\frac{x^2L_{p-2}^{2k-1}(y)+L_{p-1}^{2k}(y)}{L_p^{2k-1}(y)}
$$
is a regular function in the interval $(0,\infty )$. One obtains the
solutions of the new Schr\"o\-din\-ger equation with the help of the
transformation operator
$$
L=\left( \frac{1-4k}{2x}-\frac x2-\frac{xL_{p-1}^{2k}(y)}{L_p^{2k-1}(y)}%
-\frac d{dx}\right) (h_0+2k+2p)^{-1/2}.\eqno (19)
$$

Transformation operators (18) and (19) define the superalgebra (12) and (15)
and supercoherent states (14).

In conclusion we note that the method developed for the singular oscillator
with dynamical algebra $su(1.1)$ can be applied to other systems. We intend
to consider them in subsequent publications.

\end{document}